\documentclass[prb,twocolumn,superscriptaddress]{revtex4}
\usepackage{graphics}
\usepackage{graphicx}
\usepackage{dcolumn} 
\usepackage{bm}
\usepackage{float}
\usepackage{epsfig}
\begin{document}
\def\rhov{{\mbox{\boldmath{$\rho$}}}}
\def\tauv{{\mbox{\boldmath{$\tau$}}}}
\def\Deltav{{\mbox{\boldmath{$\Delta$}}}}
\def\Lambdav{{\mbox{\boldmath{$\Lambda$}}}}
\def\Thetav{{\mbox{\boldmath{$\Theta$}}}}
\def\Psiv{{\mbox{\boldmath{$\Psi$}}}}
\def\Phiv{{\mbox{\boldmath{$\Phi$}}}}
\def\sigmav{{\mbox{\boldmath{$\sigma$}}}}
\def\alphav{{\mbox{\boldmath{$\alpha$}}}}
\def\xiv{{\mbox{\boldmath{$\xi$}}}}
\def\oh{{\scriptsize 1 \over \scriptsize 2}}
\def\ot{{\scriptsize 1 \over \scriptsize 3}}
\def\of{{\scriptsize 1 \over \scriptsize 4}}
\def\tf{{\scriptsize 3 \over \scriptsize 4}}

\title{Classifying Possible Tilting of Oxygen Octahedra in Perovskites}

\author{A. B. Harris}

\affiliation{Department of Physics and Astronomy, University of
Pennsylvania, Philadelphia PA 19104}
\date{\today}
\begin{abstract}
The list of possible commensurate phases obtained from the parent
tetragonal phase of perovskites by
allowing the tilting of octahedra of oxygen ions is reexamined.  It
is found that many structures allowed by symmetry are not consistent with
the constraint of very rigid octahedra.
\end{abstract}
\maketitle

Many perovskite systems such as the Ruddlesden-Popper
compounds [\onlinecite{RP}]
K$_2$MgF$_4$, Ca$_3$Mn$_2$O$_7$ are constructed from layers of corner sharing
octahedra of F's or O's.  These systems exhibit many interesting technological
properties such as high $T_c$ superconductivity[\onlinecite{HTC}],
colossal magnetoresistance,[\onlinecite{CMR}]
metal insulator transitions,[\onlinecite{ARG}]
 and coupled ferroelectric and magnetic order.[\onlinecite{LOB,CF}]
Many of these properties depend sensitively on the structural
distortions from the ideal tetragonal I4/mmm structure (see Fig. 1) of space
group \#139 [space group numbering is from Ref. \onlinecite{ITC}]
which appear at structural phase
transitions.[\onlinecite{STRUC1,STRUC2,STRUC3,STRUC4,STRUC5}]
Accordingly, the accurate characterization of their structure is
essential to reach a detailed understanding of their properties
and then to fabricate new systems with enhanced desired properties.
It is not surprising then, that one of the well known
theoretical problems in crystallography is to list the possible
structures that can be obtained by cooperatively reorienting
the oxygen octahedra under the constraint of the shared vertices.
The two principal approaches to this problem
have been a) a direct enumeration of likely 
structures[\onlinecite{ALEKS}] and b) the use symmetry.[\onlinecite{HANDS}]
This last approach utilizes the celebrated computer
program[\onlinecite{COMP}] to generate isotropy subgroups tabulated in
Ref. \onlinecite{ISOTROPY}. Using this tabulation Hatch {\it et al.}
[\onlinecite{HANDS}] gave a listing for the K$_2$MgF$_4$ (214) structure
of possible irreducible representations (irreps) for distortions
by octahedral rotations. This listing was shown to be consistent with the
revised results of method a).[\onlinecite{HANDS}] This important
work has stood unchallenged for over a decade.[\onlinecite{PHASE}]
However, here we will show that some of the  listed structures
are a) counterintuitive and b) inconsistent with the fourth order term
in the Landau expansion for rigid octahedra, whose form
is less general than allowed by symmetry.

\begin{figure}
\begin{center}
\caption{(Color online) Structure of A$_3$B$_2$O$_7$ (left)
and A$_2$BF$_4$ (right). The green squares are A ions.
The B ions are at the centers of the oxygen (blue dots) octahedra.}
\includegraphics[width=6.0 cm]{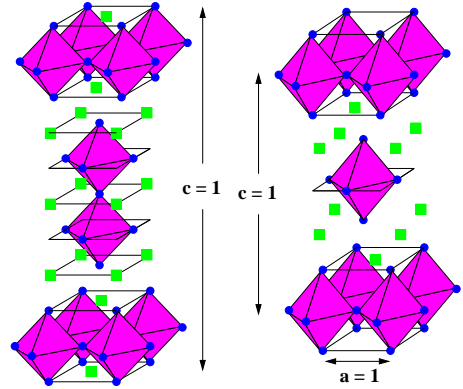}
\end{center}
\end{figure}

To see this phenomenon in its simplest guise, consider a system with
two order parameters $Q_1$ and $Q_2$ related by symmetry, for which 
the free energy assumes the form
\begin{eqnarray}
F &=& (T-T_0) [Q_1^2 + Q_2^2] + u [Q_1^2+Q_2^2]^2 + v Q_1^2Q_2^2 
\label{EQCC} \end{eqnarray}
up to fourth order in $Q$ with $u>0$.
As the temperature is lowered through the value $T_0$ the
nature of the ordering depends on the sign of $v$.  If $v$ is positive,
then ordering has either $Q_1$ or $Q_2$ zero.
If $v$ is negative, the ordering occurs with $|Q_1|=|Q_2|$.  Only at
the multicritical point[\onlinecite{MCP}] where
$v$ is zero (and also a similar sixth order anisotropy
vanishes) can one have ordering in an arbitrary direction of order
parameter space.  One may also reach such a state via a first order
transition, but for octahedral rotations this is a very unlikely
scenario, as we will explain.  However we will find that in some cases
the sign of $v$ is fixed by the intraoctahedral constraint.  In view of this
discussion it seems preferable to predict possible phases from the
form of the free energy for rigid octahedra.

As in the symmetry-based analyses, our discussion for the 214 structure
will treat only commensurate structural phase transitions
involving high symmetry wave vectors at the star of ${\bf X}=(1/2,1/2,0)$, 
of ${\bf N}=(1/2,0,1/2)$, or of ${\bf P}=(1/2,1/2,1/2)$.[\onlinecite{FN1}]
Since ${\bf X}$ is the simplest case, we will discuss it explicitly here.  

\begin{figure}[h!]
\begin{center}
\caption{\label{FIG1} (Color online) The structure of corner-sharing octahedra.
The solid (dashed) squares are the cross sections of octahedra in the plane
at $z=0$ ($z=1/2$).  For clarity the octahedra are slightly separated instead
of sharing vertices.  Here $\phi_x$ means that the $+x$ vertex moves up by an
amount $\phi_x$ and the $-x$ vertex moves down by an amount $\phi_x$
and similarly for $\phi_y$ and $\theta$ is the angle of rotation
about the $z$ axis. Also $\overline Q$ denotes $-Q$.  Left: For the star
of ${\bf X}$ and ${\bf P}$.  Right: For the star of ${\bf N}$.
For ${\bf X}$ the structure is invariant under $z \rightarrow z+1$.  For
${\bf N}$ and ${\bf P}$ the variables change sign under $z \rightarrow z+1$.}
\vspace{0.1 in}
\includegraphics[width=7.5 cm]{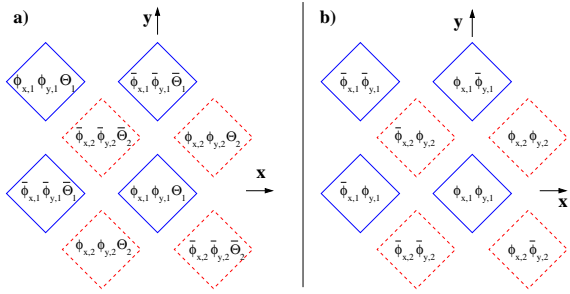}
\end{center}
\end{figure}

Instead of dealing with irreps, we will consider the most general structure
(shown in Fig. \ref{FIG1}) which can be constructed using the angular
distortions at the ${\bf X}$ wave vectors providing
that the octahedra rotate as constrained by their shared vertex.
As noted in Ref. \onlinecite{HANDS}, when one plane of octahedra are
cooperatively rotated through an angle $\phi$ by moving the shared vertices, 
the displacement of this vertex relative to what it would be if it were
not also part of an adjoining octahedron is of order $\phi^2$.  Thus
the intraoctahedral elastic energy will be quartic in the angular
variables of each plane and we introduce an expansion parameter
$\lambda \gg 1$ which is the ratio of the intraoctahedral force
constants to the other force constants of the lattice.  Since
interactions between octahedra in {\it different layers} do not
involve these large intraoctahedral force constants, there are
no interlayer couplings of order $\lambda$.  Therefore we write
the elastic free energy for the structure of Fig.  \ref{FIG1} for the star
of ${\bf X}$ as

\begin{eqnarray}
F &=& A\lambda [\theta_1^4 + \theta_2^4]
+ B\lambda [\phi_{x,1}^4 + \phi_{x,2}^4 + \phi_{y,1}^4
+ \phi_{y,2}^4] \nonumber \\ &&
+ C\lambda [\theta_1^2 (\phi_{x,1}^2 + \phi_{y,1}^2)
+ \theta_2^2 (\phi_{x,2}^2 + \phi_{y,2}^2) ]
\nonumber \\ && \
+ D\lambda [ \phi_{x,1}^2 \phi_{y,1}^2 + \phi_{x,2}^2 \phi_{y,2}^2] + F_2 \ ,
\end{eqnarray}
where $F_2$ is the free energy quadratic in the angles $\theta$ and $\phi$.
Here and below, because of the octahedral constraint quartic terms of the
form $\theta_1^2 \theta_2^2$, $\phi_{x,1}^2 \phi_{y,2}^2 + \phi_{x,2}^2
\phi_{y,1}^2$, and $\phi_{x,1}\phi_{y,1}\phi_{x,2}\phi_{y,2}$ which
are allowed by symmetry (see Table I) do not appear at order $\lambda$.
As explained below, other variables such as the displacements of 
nonoctahedral ions do not affect the symmetry of the phase we obtain.
Using Table \ref{TAB1}, we see that
the quadratic terms which are invariant under the symmetry operations are
\begin{eqnarray}
F_2 &=& \alpha [ \phi_{x,1}^2 + \phi_{y,1}^2 + \phi_{x,2}^2 + \phi_{y,2}^2]
\nonumber \\ &&
+ 2 \beta [ \phi_{x,1} \phi_{y,2} + \phi_{x,2} \phi_{y,1}]
+  \gamma [ \theta_1^2 + \theta_2^2] \ .
\end{eqnarray}
Since the angles are of order $\lambda^{-1/2}$, to evaluate $F$ to order
$1/\lambda$, we do not need to keep interoctahedral interactions of higher
than quadratic order.

\begin{table} [h!]
\caption{\label{TAB1} Effect of symmetry operations on the variables
of Fig. \ref{FIG1}. ${\cal R}_4$ is a four-fold rotation,
$m_d$ and $m_z$ are mirrors that take $x$ into $y$ and
$z$ into $-z$, respectively and $T$ is the translation (1/2,1/2,1/2).
These variables are odd under the translations $T_x=(1,0,0)$ and
$T_y=(0,1,0)$.}
\vspace{0.2 in}
\begin{tabular} {|| c | c| c| c| c||}
\hline  \hline 
& ${\cal R}_4$ & $m_d$ & $m_z$ & $T$ \\
\hline
$\phi_{x,1}$ & $\phi_{y,1}$ & $\phi_{y,1}$ & $-\phi_{x,1}$ & $\phi_{x,2}$ \\ 
$\phi_{y,1}$ & $-\phi_{x,1}$ & $\phi_{x,1}$ & $-\phi_{y,1}$ & $\phi_{y,2}$ \\ 
$\theta_1$ & $\theta_1$ & $-\theta_1$ & $\theta_1$ & $\theta_2 $ \\
$\phi_{x,2}$ & $-\phi_{y,2}$ & $\phi_{y,2}$ & $-\phi_{x,2}$ & $\phi_{x,1}$ \\ 
$\phi_{y,2}$ & $\phi_{x,2}$ & $\phi_{x,2}$ & $-\phi_{y,2}$ & $\phi_{y,1}$ \\ 
$\theta_2$ & $-\theta_2$ & $-\theta_2$ & $\theta_2$ & $\theta_1 $ \\
\hline \hline \end{tabular}
\end{table}

The structural phase transitions which we are investigating arise
when {\it one} of the channels becomes unstable, i. e. when
$\gamma$ or $\alpha - |\beta|$ passes through zero. For instance,
when only $\gamma$ becomes negative, then 
\begin{eqnarray}
\phi_{x,1}&=& \phi_{x,2}=  \phi_{y,1}= \phi_{y,2}=0 \ ,
\end{eqnarray}
so that
\begin{eqnarray}
F &=& A \lambda [ \theta_1^4 + \theta_2^4] 
- |\gamma | [ \theta_1^2 + \theta_2^2] \ ,
\end{eqnarray}
which, when minimized, leads to
\begin{eqnarray}
|\theta_1 | &=& |\theta_2| = [ - \gamma/(2A \lambda ]^{1/2} \ ,
\label{EQ7} \end{eqnarray}
which gives Cmca (64),
one of the three $\theta$-dependent structures for the star of
${\bf X}$ in Refs. \onlinecite{HANDS} and \onlinecite {PHASE}.
We do not allow the other two structures of Refs. \onlinecite{HANDS} and
\onlinecite{PHASE} which have $|\theta_1| \not= |\theta_2|$
because the octahedral constraint leads to $v=-2u$
in the language of Eq. (1).  Furthermore, the two solutions we omit are
counterintuitive. Imagine building up the structure layer by layer.
Let the first layer have $\theta=\theta_1$.  The value of $\theta$ for the
second layer is not fixed because of the frustration resulting from
the four-fold symmetry.  The sign of $\theta$ for the 
third layer is {\it not} frustrated and is $\pm \theta_1$, the sign
depending on the details of the interatomic interactions.  Thus 
$|\theta_n|=c$ and we have
two choices: either $\theta_{n+2}=\theta_n$ (ferro) or
$\theta_{n+2}=-\theta_n$ (antiferro).  The ferro (antiferro)
configuration comes from the star of ${\bf X}$ (${\bf P}$).  The relative
phase of the even and odd numbered layers is a degeneracy
similar to that in the body centered
tetragonal antiferromagnet.[\onlinecite{SHENDER,TANER}]

Now drop the $\theta$ variables, so that[\onlinecite{AXE}]
\begin{eqnarray}
F &=& B\lambda \left[ ( \phi_{x,1}^2 + \phi_{y,1}^2 )^2
+ ( \phi_{x,2}^2+\phi_{y,2}^2 )^2 \right] \nonumber \\
&& + (D-2B)\lambda
 [\phi_{x,1}^2 \phi_{y,1}^2 + \phi_{x,2}^2 \phi_{y,2}^2]
\label{EQ23} \\ &+& [(\alpha - \beta)/2] \left[
( \phi_{x,1}-\phi_{y,2})^2
+ ( \phi_{x,2}-\phi_{y,1}^2 \right]
\nonumber \\ && + [(\alpha + \beta)/2] \left[
( \phi_{x,1}+\phi_{y,2})^2
+ ( \phi_{x,2}+\phi_{y,1})^2 \right]\  .
\nonumber \end{eqnarray}
There are four directions of the ordering vector
$\Psi \equiv [\phi_{x,1}, \phi_{y,1} , \phi_{x,2} , \phi_{y,2} ]$
depending on whether or not $\alpha-\beta$ becomes critical
(negative) before $\alpha+\beta$ and whether or not $D>2B$.

\begin{figure} [h!]
\caption{\label{FIG2} (Color online)
As Fig. \ref{FIG1} for the star of ${\bf X}$ (with invariance under
$z \rightarrow z+1$).  $x$, $y$ are the tetragonal axes and
and $X$, $Y$, and $Z$ are the conventional lattice vectors after distortion.
The filled magenta circle is the tetragonal origin. $\Psi=$
$[1 \overline 1 1 \overline 1]$ for a), $[01 \overline 1 0]$ for b),
$[\overline 1 \overline 1 \overline 1 \overline 1]$ for c), and
$[0110]$ for d).  The new origin is in the $z=0$ plane, except for c)
where $z=1/4$.  Also ${\bf X} \cdot ({\bf Y} \times {\bf Z})=2$.}
\vspace{0.1 in}
\includegraphics[width=8.5 cm]{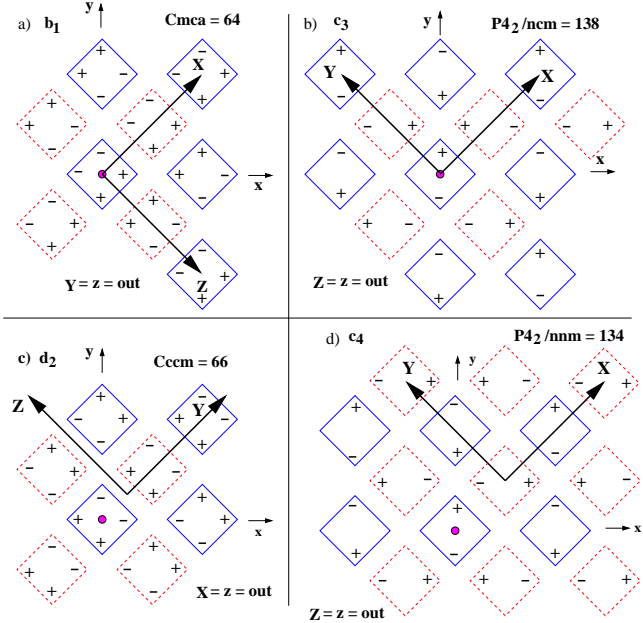}
\end{figure}

When $\alpha-\beta$ is critical and $D<2B$ then $\Psi$ is proportional to
one of $a_1 =  [11 \overline 1 \overline 1 ]$, 
$b_1 = [1  \overline 1 1 \overline 1 ]$,
$c_1 = [\overline 1 1 \overline 1 1 ]$, or
$d_1 = [\overline 1 \overline 1 1 1 ]$.
If $\alpha+\beta$ is critical and $D<2B$, then $\Psi$ is proportional to
$a_2 =  [1 1 1 1 ]$, $b_2 = [1 \overline 1 \overline 1 1 ]$,
$c_2 = [\overline 1 1 1 \overline 1 ]$,
or $d_2 = [\overline 1 \overline 1 \overline 1 \overline 1 ]$.
If $\alpha-\beta$ is critical and $D>2B$, then $\Psi$ is proportional to
$a_3 =  [1 0 0 \overline 1 ]$, $b_3 = [0  \overline 1 1 0 ]$,
$c_3 = [0 1 \overline 1 0 ]$, or $d_3 = [\overline 1 0 0 1 ]$.
If $\alpha+\beta$ is critical and $D>2B$, then $\Psi$ is proportional to
$a_4 =  [1 0 0 1 ]$, $b_4 = [0 \overline 1 \overline 1 0 ]$,
$c_4 = [0 1 1 0 ]$, or $d_4 = [\overline 1 0 0 \overline 1 ]$.
The four choices are equivalent: ${\cal R}_4 b_n = a_n$, $c_n=T_x b_n$,
and $d_n=T_x a_n$. Fig.  \ref{FIG2} shows these solutions.

Now we identify the space groups of the structures of Fig. \ref{FIG2}.
The generators of $b_1$ are $(X\pm 1/2, Y+1/2,Z)$,$(X,Y,Z+1)$,
$(\overline X, \overline Y, \overline Z)$, $(\overline X, Y, Z)$,
and $(\overline X +1/2, \overline Y, 1/2+Z)$, of
$c_4$ are $(X+1,Y,Z)$, $(X,Y+1,Z)$,
$(X,Y,Z+1)$, $(\overline X , \overline Y , \overline Z )$,
$(\overline Y + 1/2, X, Z+1/2)$, and
$(X, \overline Y + 1/2 , \overline Z + 1/2)$, of $d_2$ are
$(X \pm 1/2, Y+1/2,Z)$, $(X,Y,Z+1)$, $(\overline X , \overline Y ,\overline Z)$,
$(X,Y, \overline Z)$, and $(X, \overline Y, 1/2+Z)$, 
and of $c_3$ are $(X+ 1,Y,Z)$,
$(X,Y+1,Z)$, $(X,Y,Z+1)$,  $(\overline X, \overline Y, \overline Z)$,
$(\overline Y +1/2,X,Z+1/2)$, and $(X+1/2, \overline Y , \overline Z+1/2)$.
In comparison to Ref. \onlinecite{HANDS} we omit the structures of space
groups Pccn and Pmmm.  These structures require accessing the multicritical
point where $D=2B$.

\begin{figure} [h!]
\caption{\label{FIG3} (Color online)
As Fig. \ref{FIG2} for the star of ${\bf N}$ (with sign change under
$z \rightarrow z+1$) for a)
$[01 \overline 1 0]$, b) $[010 \overline 1]$, and
c) $[\overline 1 1 \overline 1 \overline 1]$.
$x$, $y$ are the tetragonal axes, and $X$, $Y$, and $Z$ are
$[(200)],(020),(002)]$ in a),
$[(0\overline 1 1),(100),(011)]$, in b),
and [(002),(220)(-1/2,-1/2,-1/2)] in c).}
\vspace{0.1 in}
\includegraphics[width=8.5 cm]{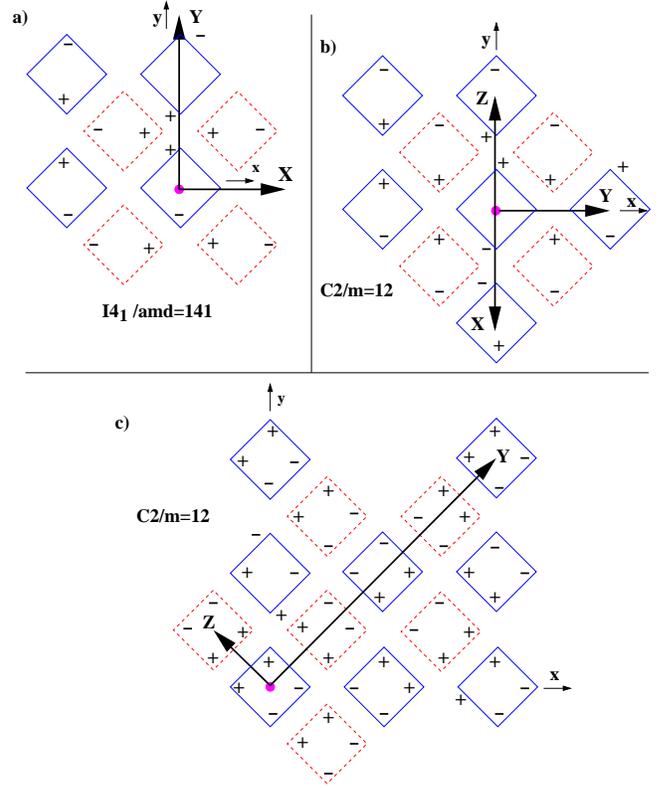}
\end{figure}

Similarly, we obtain the structure for the star of ${\bf N}$
as in Fig. 1.  The ordering vector $\Psi$ is one of
three types shown in Fig. \ref{FIG3}: $[1,1,1,1]$
which is C2/m (12), $[0101]$ which
is a different C2/m structure, or $[0110]$ which is I4$_1$/amd (141).
For the star of ${\bf N}$ we do not find the eight structures listed in
Ref. \onlinecite{HANDS} which lead to first order transitions because
these can only appear when the Landau expansion is carried to higher
order (which we discuss later). In addition, Ref. \onlinecite{HANDS}
lists two space groups Cmmm (65) for which $\Psi=[1000]$  and I4/mmm(139) 
for which $\Psi=[1100]$.  Both these are inconsistent with the
fourth order terms arising from
the rigid octahedral constraint.  They are also counterintuitive
in that they both describe states in which ordered and disordered planes
of octahedra alternate [see the discussion below Eq. (\ref{EQ7})].

\begin{figure} [h!]
\caption{\label{FIG4} (Color online)
As Fig. 3 for the star of ${\bf P}$ (variables change sign under
$z \rightarrow z+1$).  $x$, $y$ are the tetragonal axes and
and $X$, $Y$, and $Z$ are axes of the distorted structure.
In each case the new origin is in the $z=1/4$ plane.  The new
out-of-plane lattice vector has magnitude 2.}
\vspace{0.1 in}
\includegraphics[width=8.5 cm]{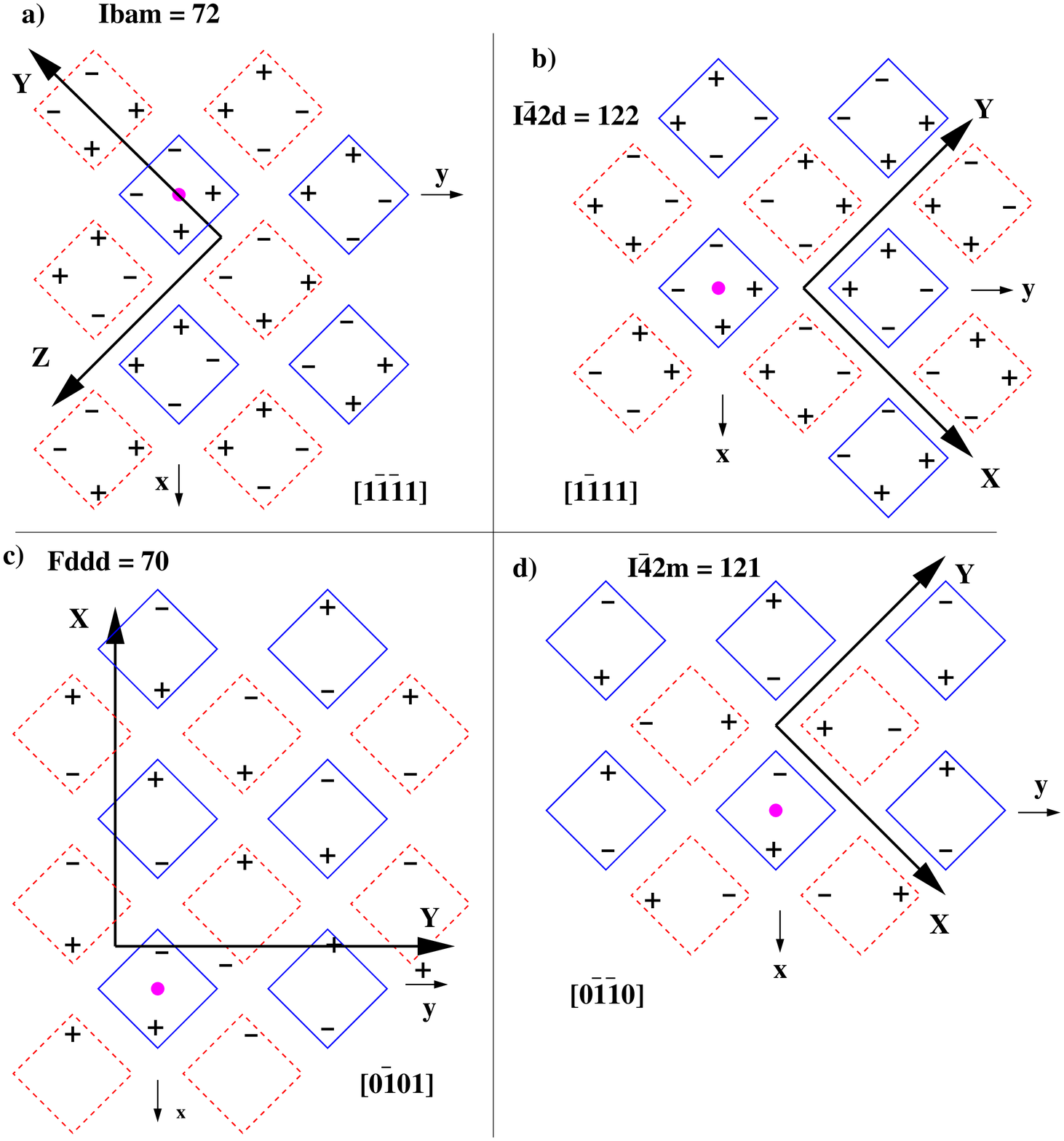}
\end{figure}

For the star of ${\bf P}$ the possible structures are those of the
left panel of Fig. 1 (with the variables changing sign under $z \rightarrow
z+1$). The only $\theta$-dependent structure has $\theta_{n+2}=-\theta_n$,
I4$_1$/acd (142),
and is one of the three structures listed in in Refs.  \onlinecite{ISOTROPY}
and \onlinecite{PHASE}. The other two structures listed there are not
admissible as explained below Eq. (\ref{EQ7}). Now consider the
$\phi$-dependent solutions.  The four allowable types of ordering vectors 
are $[1010]$, $[1001]$, $[1111]$, and $[11 \overline 1 \overline 1 ]$,
shown in Fig. \ref{FIG4}. Note that, as discussed in Ref.
\onlinecite{ISOTROPY} these $\phi$-dependent structures do not satisfy the
Lifshitz condition.  So, either the transition is (slightly) discontinuous
or the wave vector is not (exactly) equal to ${\bf P}$.  The explanation
for our omitting some of the structures found in Ref. \onlinecite{HANDS}
is the same as above.

We did not deal with the positions of the ions at the center of
the octahedra or those between the layers of octahedra.  Each such
ion sits in a stable potential well.  It is obvious that a
displacement of these ions, consistent with the symmetry
of the distorted structures we have found,
must exist.  The question is whether or not for systems without any
accidental degeneracy there is a bifurcation
so that additional space groups could be allowed when the positions
of these ``inessential" ions are taken into account.  The stable
potential well can be distorted and the placement of its minimum
will be modified by the octahedral reorientation.  But a single
minimum of a stable potential well can not be continuously deformed
into a double well without assuming an accidental vanishing of the
fourth order term in the local potential.  Similar
arguments show that the perturbative effect of the other coordinates
of the nearly rigid octahedra do not produce anomalous effects.  
It is true that in the spirit of the renormalization group the
quartic potentials we invoke can be renormalized and thereby
lead to modification, which if serious enough, could violate our
arguments.  But the stiffer the octahedra are, the less likely
such a scenario becomes. In any event, there is a regime for sufficiently
large $\lambda$ where our arguments are valid.
The results of first principle calculations[\onlinecite{PM}] indicate
that the fourth order potential used here gives a nearly perfect description
of the energy surface for octahedral rotations and, and at least
for some systems, justify our
analysis based on the Landau expansion up to quartic order.
Many of the structures of Ref. \onlinecite{HANDS} which we do not accept
are those which arise from discontinuous transitions caused by higher than
quartic terms in the free energy (which we omit).  But several structures
we omit (such as those with $|\theta_1| \not= |\theta_2|$ or with
disordered sublattices) are omitted
because of the special form arising from the intraoctahedral constraint.

Experimentally, it is striking that the structures observed as distortions
from the tetragonal phase are in our much shorter list.  For instance,
in the data cited on p 313 and ff of Ref. \onlinecite{PHASE} five systems
with $\phi$ tilts are shown which go into either Cmca (64) or 
P4$_2$/ncm (138), except for Rb$_2$CdCl$_4$ whose structure is
uncertain: either Cmca or Fccm (which is on neither our list
nor that of Ref. \onlinecite{HANDS} because it involves two irreps).
Systems (other than Rb$_2$CdCl$_4$ subseqently discussed in Ref.
\onlinecite{PHASE}) in Table III of Ref. \onlinecite{HANDS} likewise
go into either Cmca or P4$_2$/ncm.

To summarize: we find that the rigid
octahedral constraint eliminates all the structures in Table I of Ref.
\onlinecite{HANDS} for which the octahedral tilting transitions are
discontinuous and, in addition, those that are allowed by symmetry to
be continuous but which involve disordered sublattices.
Elsewhere we will give the results of our approach to encompass
sequential phase transitions which involve two distinct irreps.

I would like to thank the following for helpful correspondence:
M. Perez-Mato, C. J. Fennie, B. Campbell, H. T. Stokes, and M. V. Lobanov.


\begin{thebibliography} {99}
\bibitem{RP}
S. N. Ruddlesden and P. Popper, Acta Cryst. {\bf 11}, 54 (1958).
\bibitem{HTC}
J. D. Bednorz and K. A. M\"uller,  Z. Phys. {\bf B64}, 189 (1986).
\bibitem{CMR} 
Y. Tokura (Ed.), {\it Colossal Magnetoresistance Oxides}, Monograph
in Condensed Matter Science, Gordon and Breach, London, 2000.
\bibitem{ARG} 
J. F. Mitchell {\it et al.},
J. Phys. Chem. B {\bf 105}, 10732 (2001).
\bibitem{LOB} 
M. V. Lobanov {\it et al.},
J. Phys.: Condens. Matter {\bf 16}, 5339 (2004).
\bibitem{CF}
N. A. Benedek and C. J. Fennie, arXiv: 1007.1003.
\bibitem{ITC}
A. J. C. Wilson, {\it International Tables for Crystallography}
(Kluwer Academic, Dordrecht, 1995), Vol. A.
\bibitem{STRUC1} 
K. R. Poepplemeier {\it et al.},
J. Solid State Chem. {\bf 45}, 71 (1982).
\bibitem{STRUC2} 
M. E Leonowicz, K. R. Poepplemeier, and J. M. Longo, J. Solid State
Chem. {\bf 71}, 59 (1985).
\bibitem{STRUC3} 
P. D. Battle {\it et al},
Chem. Mater. {\bf 9}, 552 (1997).
\bibitem{STRUC4} 
P. D. Battle {\it et al.}.
Chem. Mater. {\bf 10}, 658 (1998).
\bibitem{STRUC5} 
I. D. Fawcett {\it et al.},
Chem. Mater. {\bf 10}, 3643 (1998).
\bibitem{ALEKS}
K. S. Aleksandrov, B. V. Beznosikov, and S. V. Misyul, Phys.
Status Solidi A {\bf 104}, 529 (1987).
\bibitem{HANDS}
D. M. Hatch, H. T. Stokes, K. Aleksandrov, and S. V. Misyul,
Phys. Rev. B {\bf 39} 9282 (1989).
\bibitem{COMP}
ISODISTORT (accessible from the internet).
\bibitem{ISOTROPY}
H. T. Stokes and D. M. Hatch, {\it Isotropy Subgroups of the 230 
Crystallographic Space Groups} (World-Scientific, Singapore, 1988).
\bibitem{PHASE}
K. S. Aleksandrov and J. Bartolom\'e, Phase Transitions {\bf 74}, 255
(2001).
\bibitem{FN1}
Lengths are in units of lattice constants $a_i$ and wave
vectors in units of $2 \pi /a_i$.
\bibitem{MCP}
See Wikipedia under ``Multicritical point."
\bibitem{PM}
J. M. Perez-Mato {\it et al.},
Phys. Rev. B {\bf 70}, 214111 (2004).
\bibitem{SHENDER}
E. F. Shender, Sov. Phys. JETP {\bf 56}, 178 (1982).
\bibitem{TANER}
T. Yildirim, A. B. Harris, and E. F. Shender, Phys. Rev. B {\bf 53},
6455 (1996).
\bibitem{AXE}
J. D. Axe {\it et al.}, Phys. Rev. Lett. {\bf 62}, 2751 (1989).
The Landau free energy in this reference is equivalent to what is written
here for the case when $\phi_{x,1}=-\phi_{y,2}=Q_1$ and 
$\phi_{y,1}=-\phi_{x,2}=Q_2$.
\end{thebibliography}
\end{document}